\documentclass[12pt]{article}
\begin{document}

\newtheorem{theo}{Theorem}[section]
\newtheorem{defi}[theo]{Definition}
\newtheorem{prop}[theo]{Proposition}
\newtheorem{corr}[theo]{Corollary}
\newtheorem{lemm}[theo]{Lemma}
\newtheorem{exam}[theo]{Example}

\newcommand{\Eh}{\ensuremath{\hat{E}^{t}}}
\newcommand{\Ep}{\ensuremath{E^{t}_{+}}}
\newcommand{\Et}{\ensuremath{\tilde{E}^{t}}}
\newcommand{\Hp}{\ensuremath{\mathbf{H}_{+}}}
\newcommand{\Lp}{\ensuremath{L_{p}(\Re_{+})}}
\newcommand{\mt}{\ensuremath{\mathbf{\hat{t}}}}
\newcommand{\U}{\ensuremath{U(\theta,\tau)}}

\title{Feynman's Path Integrals \\
        as Evolutionary Semigroups}

\author{David W. Dreisigmeyer and Peter M. Young\\
        Electrical and Computer Engineering\\
        Colorado State University}

\maketitle

\begin{abstract}

We show that, for a class of systems described by a Lagrangian
\begin{eqnarray*}
    L(x,\dot{x},t) & = & \frac{1}{2}\dot{x}^{2} - V(x,t)
\end{eqnarray*}
the propagator
\begin{eqnarray*}
    K\left(x^{''},t^{''};x^{'},t^{'}\right) & = &
    \int
        e^{\frac{i}{\hbar}\int_{t^{'}}^{t^{''}}dt
        L \left( x,\dot{x},t \right)}\mathcal{D}\left[ x(t) \right]
\end{eqnarray*}
can be reduced via Noether's Theorem to a standard path integral
multiplied by a phase factor.  Using Henstock's integration
technique, this path integral is given a firm mathematical basis.
Finally, we recast the propagator as an evolutionary semigroup.

\end{abstract}

\section{Noether Invariants and Path Integrals}
Here we deal with systems described by a Lagrangian of the form
($m \equiv 1$)
\begin{eqnarray}
\label{1}
    L(x,\dot{x},t) & = & \frac{1}{2}\dot{x}^{2} - V(x,t)
\end{eqnarray}
The propagator for this system is given by the path integral
\begin{eqnarray}
\label{2}
    K\left(x^{''},t^{''};x^{'},t^{'}\right) & = &
    \int
        e^{\frac{i}{\hbar}\int_{t^{'}}^{t^{''}}dt
        L \left( x,\dot{x},t \right)}\mathcal{D}\left[ x(t) \right]
\end{eqnarray}
We will use Noether invariants to help simplify (\ref{2}). Our
goal is to remove the time-dependence from the integral in the
exponential.  The way this will be accomplished is through global
time and scale transformations.

Suppose the transformations (assumed to be sufficiently smooth in
what follows)
\begin{eqnarray}
\label{3}
    x & \rightarrow & \hat{x}  = x + \epsilon\chi(x,t)
\end{eqnarray}
\begin{eqnarray}
\label{4}
    t & \rightarrow & \hat{t}  =  t + \epsilon\tau(x,t)
\end{eqnarray}
leave the action
\begin{eqnarray*}
\label{5}
    S[x] & = & \int_{t^{'}}^{t^{''}} L dt
\end{eqnarray*}
invariant (for particular functions $\chi$ and $\tau$).  That is,
there exists a function $\Omega(x,t)$ such that, for every $x(t)$
we have (see for example~\cite{SaCa})
\begin{eqnarray}
\label{6}
    \int_{\hat{t}^{'}}^{\hat{t}^{''}}
    \hat{L}(\hat{x},\frac{d\hat{x}}{d\hat{t}},\hat{t})d\hat{t} & = &
    \int_{t^{'}}^{t^{''}} L(x,\dot{x},t)dt
    + \epsilon\int_{t^{'}}^{t^{''}}
    \frac{d\Omega}{dt}(x,t)dt
\end{eqnarray}
where we have ignored terms of $O(\epsilon^{2})$ and higher.  Note
that the integral involving $\dot{\Omega}$ adds only a constant
term since $x(t^{'})$ and $x(t^{''})$ are fixed. Also, if
$\hat{L}$ has the same form as $L$ then $\Omega \equiv 0$.  We
also assume that (\ref{3}) and (\ref{4}) are invertible in the
sense that there are functions $\hat{\chi}$ and $\hat{\tau}$ (also
assumed to be sufficiently smooth) such that
\begin{eqnarray}
\label{3a}
    x & = & \hat{x} + \epsilon\hat{\chi}(\hat{x},\hat{t})
\end{eqnarray}
\begin{eqnarray}
\label{4a}
    t & = & \hat{t} + \epsilon\hat{\tau}(\hat{x},\hat{t})
\end{eqnarray}

For (\ref{5}) to hold it is necessary and sufficient to have
\cite{LoRu}
\begin{eqnarray}
\label{6}
    \hat{L}(\hat{x},\frac{d\hat{x}}{d\hat{t}},\hat{t})\frac{d\hat{t}}{dt} & = &
    L(x,\dot{x},t)
    +
    \epsilon\frac{d\Omega}{dt}(x,t)
\end{eqnarray}
Now, (\ref{6}) must hold for all infinitesimal values of
$\epsilon$ in (\ref{3}) and (\ref{4}). So, if we differentiate
both sides of (\ref{6}) with respect to $\epsilon$, remember that
$L$ does not depend on $\epsilon$, and set $\epsilon = 0$ we have
\begin{eqnarray}
\label{7}
    \left. \left(\frac{d}{d\epsilon} \hat{L}\right)\frac{d\hat{t}}{dt}
    \right|_{\epsilon = 0} + \hat{L}
    \left. \left(\frac{d}{d\epsilon} \frac{d\hat{t}}{dt}\right)
    \right|_{\epsilon = 0} & = & \frac{d\Omega}{dt}
\end{eqnarray}
At $\epsilon = 0$ we have $\hat{x} = x$, $d\hat{x}/d\hat{t} =
\dot{x}$, $\hat{t} = t$ and $\hat{L} = L$.  Also,
\begin{eqnarray}
\label{8}
    \left. \frac{d}{d\epsilon}\frac{d\hat{t}}{dt}
    \right|_{\epsilon= 0} & = & \frac{d}{d\epsilon}\left. \left(1
    + \epsilon \frac{\partial \tau}{\partial x} \dot{x} + \epsilon
    \frac{\partial \tau}{\partial t}\right)\right|_{\epsilon =
    0}      \\
    & = & \left. \left(\frac{\partial \tau}{\partial x} \dot{x} +
    \frac{\partial \tau}{\partial t}\right) \right|_{\epsilon =
    0}    \nonumber  \\
    & = & \frac{\partial \tau}{\partial t} + \frac{\partial \tau}{\partial x}
    \dot{x} \nonumber   \\
    & = & \frac{d\tau}{dt} \nonumber
\end{eqnarray}
So,
\begin{eqnarray*}
    \hat{L}\left. \left(\frac{d}{d\epsilon} \frac{d\hat{t}}{dt}\right)
    \right|_{\epsilon = 0} & = & L\dot{\tau}
\end{eqnarray*}
Also notice from (\ref{8}) that
\begin{eqnarray}
\label{9}
    \left. \frac{d\hat{t}}{dt} \right|_{\epsilon = 0} & = & 1
\end{eqnarray}

Returning to (\ref{7}) we have
\begin{eqnarray*}
    \left. \frac{d}{d\epsilon} \hat{L}\right|_{\epsilon = 0} & = &
    \left.
    \left(\frac{\partial\hat{L}}{\partial\hat{x}}\frac{\partial\hat{x}}{\partial\epsilon}
    + \frac{\partial\hat{L}}{\partial(d\hat{x}/d\hat{t})}\frac{\partial(d\hat{x}/d\hat{t})}{\partial\epsilon}
    +
    \frac{\partial\hat{L}}{\partial\hat{t}}\frac{\partial\hat{t}}{\partial\epsilon}\right)
    \right|_{\epsilon = 0}
\end{eqnarray*}
Now, from (\ref{6}) and using (\ref{3a})
\begin{eqnarray}
\label{10}
    \left. \frac{\partial\hat{L}}{\partial\hat{x}}
    \right|_{\epsilon = 0} & = & \left. \frac{\partial L}{\partial\hat{x}}
    \right|_{\epsilon = 0} + \left.\epsilon \frac{\partial \Omega}{\partial\hat{x}}
    \right|_{\epsilon = 0}  \\
    & = & \frac{\partial L}{\partial x}
    \left. \frac{\partial x}{\partial\hat{x}}
    \right|_{\epsilon = 0} + \frac{\partial L}{\partial \dot{x}}
    \left. \frac{\partial \dot{x}}{\partial\hat{x}}
    \right|_{\epsilon = 0} + \epsilon \frac{\partial \Omega}{\partial x}
    \left. \frac{\partial x}{\partial\hat{x}}
    \right|_{\epsilon = 0} \nonumber \\
    & = & \left. \frac{\partial L}{\partial x}
    \right|_{\epsilon = 0}  \nonumber \\
    & = & \frac{\partial L}{\partial x} \nonumber
\end{eqnarray}
Similarly, we have
\begin{eqnarray}
\label{11}
    \left. \frac{\partial\hat{L}}{\partial(d\hat{x}/d\hat{t})}
    \right|_{\epsilon = 0} & = &  \frac{\partial L}{\partial\dot{x}}
\end{eqnarray}
\begin{eqnarray}
\label{12}
    \left. \frac{\partial\hat{L}}{\partial\hat{t}}
    \right|_{\epsilon = 0} & = & \frac{\partial L}{\partial t}
\end{eqnarray}
Looking at (\ref{3}) and (\ref{4}) we see that
\begin{eqnarray}
\label{13}
    \left. \frac{\partial\hat{x}}{\partial \epsilon}
    \right|_{\epsilon = 0} & = & \chi
\end{eqnarray}
\begin{eqnarray}
\label{14}
    \left. \frac{\partial \hat{t}}{\partial \epsilon}
    \right|_{\epsilon = 0} & = & \tau
\end{eqnarray}
So now using (\ref{9}) to (\ref{14})
\begin{eqnarray*}
    \left. \left( \frac{d}{d\epsilon} \hat{L} \right)
    \frac{d\hat{t}}{dt} \right|_{\epsilon = 0} & = &
    \left(\frac{\partial L}{\partial x} \chi + \frac{\partial L}{\partial
    \dot{x}}\left. \left( \frac{\partial (d\hat{x}/d \hat{t})}{\partial
    \epsilon}\right) \right|_{\epsilon = 0} + \frac{\partial L}{\partial
    t}\tau \right)
\end{eqnarray*}
We only need to find
\begin{eqnarray*}
    \left. \frac{\partial (d\hat{x}/d \hat{t})}{\partial
    \epsilon}\right|_{\epsilon = 0}
\end{eqnarray*}
Notice that
\begin{eqnarray}
\label{15}
    \frac{d\hat{x}}{d\hat{t}}\frac{d\hat{t}}{dt} & = &
    \frac{d\hat{x}}{dt}
\end{eqnarray}
So
\begin{eqnarray}
\label{16}
    \left(\frac{\partial}{\partial\epsilon}\frac{d\hat{x}}{d\hat{t}}\right)\frac{d\hat{t}}{dt}
    + \frac{d\hat{x}}{d\hat{t}}\left(\frac{\partial}{\partial\epsilon}\frac{d\hat{t}}{dt}\right)
    & = & \frac{\partial}{\partial\epsilon}\frac{d\hat{x}}{dt}
\end{eqnarray}
Evaluating (\ref{16}) at $\epsilon = 0$ and using (\ref{8}) and
(\ref{9}) gives us
\begin{eqnarray}
\label{17}
    \left.
    \frac{\partial}{\partial\epsilon}\frac{d\hat{x}}{d\hat{t}}\right|_{\epsilon
    = 0} + \left. \frac{d\hat{x}}{d\hat{t}} \right|_{\epsilon = 0}
    \dot{\tau} & = & \left.
    \frac{\partial}{\partial\epsilon}\frac{d\hat{x}}{dt}\right|_{\epsilon
    = 0}
\end{eqnarray}
Using (\ref{3}) in (\ref{15}) we see that
\begin{eqnarray}
\label{18}
    \left. \frac{d\hat{x}}{d\hat{t}} \right|_{\epsilon = 0} & = &
    \dot{x}
\end{eqnarray}
Also, from (\ref{3})
\begin{eqnarray}
\label{19}
    \frac{d\hat{x}}{dt} & = & \dot{x} + \epsilon
    \frac{\partial\chi}{\partial x}\dot{x} +
    \epsilon\frac{\partial\chi}{\partial t}
\end{eqnarray}
Using (\ref{18}) and (\ref{19}) in (\ref{17}) gives us
\begin{eqnarray*}
    \left.
    \frac{\partial}{\partial\epsilon}\frac{d\hat{x}}{d\hat{t}}
    \right|_{\epsilon = 0} & = & \left.
    \frac{\partial}{\partial\epsilon} \left(\dot{x} + \epsilon
    \frac{\partial\chi}{\partial x}\dot{x} +
    \epsilon\frac{\partial\chi}{\partial
    t}\right)\right|_{\epsilon= 0} - \dot{x}\dot{\tau}  \\
    & = & \frac{\partial\chi}{\partial x}\dot{x} + \frac{\partial\chi}{\partial
    t} - \dot{x}\dot{\tau}  \\
    & = & \frac{d\chi}{dt} - \dot{x}\dot{\tau}  \\
    & = & \dot{\chi} - \dot{x}\dot{\tau}
\end{eqnarray*}
So now we have
\begin{eqnarray*}
    \left. \left(
    \frac{d}{d\epsilon}\hat{L}\right)\frac{d\hat{t}}{dt}\right|_{\epsilon
    = 0} & = & \frac{\partial L}{\partial x}\chi + \frac{\partial L}{\partial
    \dot{x}}(\dot{\chi} - \dot{x}\dot{\tau}) + \frac{\partial L}{\partial
    t}\tau
\end{eqnarray*}
Then (\ref{7}) becomes
\begin{eqnarray}
\label{20}
    \frac{\partial L}{\partial x}\chi + \frac{\partial L}{\partial
    \dot{x}}(\dot{\chi} - \dot{x}\dot{\tau}) + \frac{\partial L}{\partial
    t}\tau + L\dot{\tau} & = & \frac{d\Omega}{dt}
\end{eqnarray}

Now
\begin{eqnarray*}
    \frac{\partial L}{\partial \dot{x}}\dot{\chi} & = &
    \frac{d}{dt}\left(\frac{\partial L}{\partial
    \dot{x}}\chi\right) - \frac{d}{dt}\left( \frac{\partial L}{\partial
    \dot{x}}\right)\chi
\end{eqnarray*}
Then (\ref{20}) becomes
\begin{eqnarray*}
    \frac{\partial L}{\partial x}\chi +L\dot{\tau} +\frac{d L}{dt}\tau
    - \frac{\partial L}{\partial x}\dot{x}\tau +
    \frac{d}{dt}\left(\frac{\partial L}{\partial \dot{x}}\chi\right) -
    \left(\frac{d}{dt}\frac{\partial L}{\partial \dot{x}}\right) \chi
    \\
    - \left(\frac{\partial L}{\partial \dot{x}}\dot{x}\dot{\tau} +
    \frac{\partial L}{\partial \dot{x}}\ddot{x}\tau\right) & = &
    \frac{d\Omega}{dt}
\end{eqnarray*}
Using
\begin{eqnarray*}
    \frac{\partial L}{\partial \dot{x}}\dot{x}\dot{\tau} + \frac{\partial L}{\partial
    \dot{x}}\ddot{x}\tau & = & \frac{d}{dt}\left(\frac{\partial L}{\partial
    \dot{x}}\dot{x}\tau\right) - \left(\frac{d}{dt}\frac{\partial L}{\partial
    \dot{x}}\right)\dot{x}\tau
\end{eqnarray*}
gives
\begin{eqnarray*}
    \frac{\partial L}{\partial
    x}\chi - \left(\frac{d}{dt}\frac{\partial L}{\partial
    \dot{x}}\right)\chi + L\dot{\tau} + \frac{d
    L}{dt}\tau - \frac{\partial L}{\partial
    x}\dot{x}\tau + \frac{d}{dt}\left(\frac{\partial L}{\partial
    \dot{x}}\chi\right)     \\
    - \frac{d}{dt} \left(\frac{\partial
    L}{\partial \dot{x}}\dot{x}\tau\right) +
    \left(\frac{d}{dt}\frac{\partial L}{\partial
    \dot{x}}\right)\dot{x}\tau & = & \frac{d\Omega}{dt}
\end{eqnarray*}
Rearranging some we have
\begin{eqnarray}
\label{21}
    \left[\frac{\partial L}{\partial x} - \frac{d}{dt}\left(\frac{\partial L}{\partial
    \dot{x}}\right)\right](\chi - \dot{x}\tau) +
    \frac{d}{dt}\left[L\tau + \frac{\partial L}{\partial
    \dot{x}}(\chi - \dot{x}\tau)\right] & = & \frac{d\Omega}{dt}
\end{eqnarray}

So, this gives us
\begin{eqnarray}
\label{22}
    (\chi - \dot{x}\tau)\left[\frac{\partial L}{\partial
    x}-\frac{d}{dt}\left(\frac{\partial L}{\partial
    \dot{x}}\right)\right] & = & \frac{d}{dt}I(x,\dot{x},t)
\end{eqnarray}
where
\begin{eqnarray}
\label{23}
    I(x,\dot{x},t) & = & -L\tau - \frac{\partial L}{\partial
    \dot{x}}(\chi - \dot{x}\tau) + \Omega
\end{eqnarray}
\textit{Hamilton's principle} states that the actual path followed
will be an extreme value of the action integral.  That is
\begin{eqnarray*}
    \delta S[x] & = & \delta \int_{t^{'}}^{t^{''}} L dt \\
                & = & 0
\end{eqnarray*}
This yields the Euler-Lagrange equation of motion
\begin{eqnarray}
\label{24}
    \frac{d}{dt}\left(\frac{\partial L}{\partial \dot{x}}\right) -
    \frac{\partial L}{\partial x} & = & 0
\end{eqnarray}
Using (\ref{24}) in (\ref{22}) we have
\begin{theo}[Noether's First Theorem (for one parameter)]
    Along a path $x(t)$ satisfying (\ref{24}), $I(x,\dot{x},t)$ as
    given in (\ref{23}) is constant.
\end{theo}

Now we put our Lagrangian (\ref{1}) into (\ref{23}) to get
\begin{eqnarray}
\label{25}
    I(x,\dot{x},t) & = & \frac{1}{2}\dot{x}^{2}\tau - \dot{x}\chi
    +V\tau + \Omega
\end{eqnarray}
So a Lagrangian of the form
\begin{eqnarray*}
    L(x,\dot{x},t) & = & \frac{1}{2}\dot{x}^{2} - V(x,t)
\end{eqnarray*}
leads to an invariant quadratic in $\dot{x}$.  We ask, is there a
restriction on the form of $V$ that will allow an invariant $I$?
The answer is yes.

In order to determine the form of $V$ we follow the method
developed by Lewis and Leach (see \cite{LeLe4}, \cite{LeLe3},
\cite{LeLe2}, \cite{LeLe1} and \cite{Ray}).  Since we know our
invariant is quadratic in $\dot{x}$, we start with a general
invariant
\begin{eqnarray}
\label{25a}
    I(x,\dot{x},t) & = & f_{2}(x,t)\dot{x}^{2} + f_{1}(x,t)\dot{x}
    + f_{0}(x,t)
\end{eqnarray}
where $f_{0}$, $f_{1}$ and $f_{2}$ are functions of $x$ and $t$
satisfying the condition
\begin{eqnarray*}
    \frac{dI}{dt} & = & \frac{\partial I}{\partial t} + \frac{\partial I}{\partial
    x}\dot{x} + \frac{\partial I}{\partial \dot{x}}\ddot{x} \\
    & = & \frac{\partial I}{\partial t} + \frac{\partial I}{\partial
    x}\dot{x} -\frac{\partial I}{\partial \dot{x}}\frac{\partial V}{\partial x}
    \\
    & = & 0
\end{eqnarray*}
We used (\ref{24}) in going to the second equality above.  For
(\ref{25a})
\begin{eqnarray*}
    \frac{\partial I}{\partial t} & = & f_{2t}\dot{x}^{2} +
    f_{1t}\dot{x} + f_{0t}  \\
    \frac{\partial I}{\partial x}\dot{x} & = & f_{2x}\dot{x}^{3} +
    f_{1x}\dot{x}^{2} + f_{0x}\dot{x}   \\
    -\frac{\partial I}{\partial \dot{x}}\frac{\partial V}{\partial
    x} & = & -2f_{2}\dot{x}\frac{\partial V}{\partial x} - f_{1}\frac{\partial V}{\partial x}
\end{eqnarray*}
where the additional subscripts denote partial differentiation
with respect to $t$ or $x$.

So now we have
\begin{eqnarray}
\label{26}
    (f_{2x})\dot{x}^{3} + (f_{2t} + f_{1x})\dot{x}^{2} + (-2f_{2}\frac{\partial V}{\partial
    x} + f_{1t} + f_{0x})\dot{x}    \\
     + (-f_{1}\frac{\partial V}{\partial
    x} + f_{0t}) & = & 0 \nonumber
\end{eqnarray}
Since $f_{0}$, $f_{1}$ and $f_{2}$ are independent of $\dot{x}$,
we require each of the coefficients in (\ref{26}) to vanish
separately.  This gives us
\begin{eqnarray*}
    f_{2x} & = & 0
\end{eqnarray*}
so
\begin{eqnarray*}
    f_{2}(t) & = & 2\alpha(t)
\end{eqnarray*}
Then,
\begin{eqnarray*}
    f_{1x} & = & -2\dot{\alpha}
\end{eqnarray*}
which leads to
\begin{eqnarray*}
    f_{1}(x,t) & = & -2\dot{\alpha}(t)x + \beta(t)
\end{eqnarray*}
Then we have
\begin{eqnarray}
\label{27}
    -4\alpha\frac{\partial V}{\partial x} - 2\ddot{\alpha}x +
    \dot{\beta} + f_{0x} & = & 0
\end{eqnarray}
\begin{eqnarray}
\label{28}
    (2\dot{\alpha}x - \beta)\frac{\partial V}{\partial x} + f_{0t}
    & = & 0
\end{eqnarray}
Integrating (\ref{27}) with respect to $x$ gives us
\begin{eqnarray*}
    -4\alpha\left[V + \gamma_{1}(t)\right] -
    \ddot{\alpha}x^{2} + \dot{\beta}x + \left[f_{0} +
    \gamma_{2}(t)\right] & = & \gamma_{3}(t)
\end{eqnarray*}
or
\begin{eqnarray}
\label{28a}
    V(x,t) & = & \frac{1}{4\alpha}(f_{0} - \ddot{\alpha}x^{2} +
    \dot{\beta} x) + \gamma(t)
\end{eqnarray}
Using $V$ in (\ref{28}) and solving the PDE  for $f_{0}$ via the
method of characteristics results in
\begin{eqnarray*}
    f_{0}(x,t) & = & G\left(\frac{x}{\alpha^{1/2}} +
    \frac{1}{4}\int^{t}\frac{\beta(t^{*})}{\alpha^{3/2}(t^{*})}dt^{*}\right)
    + \frac{1}{2\alpha}(\dot{\alpha}x - \frac{1}{2}\beta)^{2}
\end{eqnarray*}
where $G(\cdot)$ is an arbitrary function of its argument.

If we define $\rho(t)$ and $a(t)$ by
\begin{eqnarray*}
    \rho(t) & = & 2\alpha(t)^{1/2}  \\
    -\frac{a(t)}{\rho(t)} & = & \frac{1}{8}\int^{t}\frac{\beta(t^{*})}
    {\alpha^{3/2}(t^{*})}dt^{*}
\end{eqnarray*}
and use the solution for $f_{0}$ in (\ref{28a}), we have that the
potential is
\begin{eqnarray*}
    V(x,t) & = & \left( \frac{\ddot{\rho}a - \rho\ddot{a}}{\rho}
    \right)x - \frac{1}{2}
    \frac{\ddot{\rho}}{\rho}x^{2} + \frac{1}{\rho^{2}}F\left(
    \frac{x-a}{\rho}\right)
\end{eqnarray*}
where $F(u) \doteq G(2u)$.  We have removed an irrelevant function
of $t$ from $V$ since it only adds a constant to the value of the
action $S[x]$. This potential has the associated Noether invariant
\begin{eqnarray*}
    I(x,\dot{x},t) & = & \frac{1}{2}\left[ \rho(\dot{x} -
    \dot{a}) - \dot{\rho}(x - a) \right]^{2} + F\left(
    \frac{x - a}{\rho} \right)
\end{eqnarray*}
Finding the parts of $I$ that depend on $\dot{x}$, and
$\dot{x}^{2}$
\begin{eqnarray*}
    I & = & \frac{\rho^{2}}{2}\dot{x}^{2} -
    \rho\left[\rho\dot{a} + \dot{\rho}(x - a)\right]
    \dot{x} + \cdots
\end{eqnarray*}
and comparing with (\ref{25}) leads to
\begin{eqnarray*}
    \tau & = & \rho^{2} \\
    \chi & = & \rho\left[\rho\dot{a} + \dot{\rho}(x -
    a)\right]
\end{eqnarray*}

In summary, we can start with any two functions $a(t)$ and
$\rho(t)$ that have continuous second derivatives on our time
period of interest (since we will want V to be continuous below).
With these we define a Lagrangian of the form
\begin{eqnarray*}
    L(x,\dot{x},t) & = & \frac{1}{2}\dot{x}^{2} - V(x,t)
\end{eqnarray*}
where
\begin{eqnarray}
\label{29}
    V(x,t) & = & \left( \frac{\ddot{\rho}a - \rho\ddot{a}}{\rho}
    \right)x - \frac{1}{2}
    \frac{\ddot{\rho}}{\rho}x^{2} + \frac{1}{\rho^{2}}F\left(
    \frac{x-a}{\rho}\right)
\end{eqnarray}
and $F(\cdot)$ is an arbitrary function of its argument.  The
action
\begin{eqnarray*}
    S[x] & = & \int_{t^{'}}^{t^{''}} L dt
\end{eqnarray*}
will be invariant under the transformations
\begin{eqnarray*}
    x & \rightarrow & \hat{x}(x,t)  =  x +
    \epsilon\rho\left[\rho\dot{a} + \dot{\rho}(x -
    a)\right]  \\
    t & \rightarrow & \hat{t}(t)  =  t + \epsilon\rho^{2}(t)
\end{eqnarray*}
Associated with this system will be the Noether invariant
\begin{eqnarray}
\label{30}
    I(x,\dot{x},t) & = & \frac{1}{2}\left[ \rho(\dot{x} -
    \dot{a}) - \dot{\rho}(x - a) \right]^{2} + F\left(
    \frac{x - a}{\rho} \right)
\end{eqnarray}
Note that $V$ can have an arbitrary additive function of $t$
\begin{eqnarray*}
    V & \rightarrow & V + g(t)
\end{eqnarray*}
This only adds a constant to the action so we can set $g(t) \equiv
0$.
\begin{exam}[The 1-D Time-Dependent Harmonic Oscillator]
\label{TDHOexam}
Consider the case where
\[
\begin{array}{ll}
    C\rho(t) = a(t) & 0\leq C<\infty
\end{array}
\]
and $F \equiv 0$.  Then we have
\begin{eqnarray*}
    V(x,t) & = & -\frac{1}{2} \frac{\ddot{\rho}}{\rho}x^{2} \\
            & = & \frac{1}{2}\omega(t)x^{2}
\end{eqnarray*}
where $\omega(t) \doteq -\ddot{\rho}/\rho$.  This is the harmonic
oscillator with a time-dependent frequency.  Now, we see that the
Lagrangian
\begin{eqnarray*}
    L(x,\dot{x},t) & = & \frac{1}{2}\dot{x}^{2}
    -\frac{1}{2}\omega(t)x^{2}
\end{eqnarray*}
will be invariant under the transformations
\begin{eqnarray*}
    \hat{x} & = & x + \frac{\epsilon}{2}\left(\frac{d}{dt}\rho^{2}\right)x    \\
    \hat{t} & = & t + \epsilon \rho^{2}
\end{eqnarray*}
The Noether invariant of this system is given by
\begin{eqnarray*}
    I(x,\dot{x},t) & = & \frac{1}{2}[\rho\dot{x} -
    \dot{\rho}x]^{2}
\end{eqnarray*}
The reader is invited to show what happens when we have $C = 0$
above and
\[
\begin{array}{ll}
    F(t) = \frac{1}{2} K \left( \frac{x}{\rho} \right)^{2} & 0 < K
    < \infty
\end{array}
\]
\end{exam}

Now that we have an expression for $V$, we put this into (\ref{1})
to get (see \cite{KhLa1})
\begin{eqnarray*}
    L(x,\dot{x},t) & = & \frac{d}{dt}X + L_{0}
\end{eqnarray*}
where
\begin{eqnarray*}
    L_{0}(x,\dot{x},t) & = &
        \frac{\rho^{2}}{2}\left[\frac{d}{dt}\left(\frac{x -
        a}{\rho}\right)\right]^{2} - \frac{1}{\rho^{2}}
        F\left(\frac{x - a}{\rho}\right)
\end{eqnarray*}
and
\begin{eqnarray*}
    X(t) & = & \frac{\dot{\rho}}{2\rho}x^{2} + \frac{W}{\rho}x -
        G\\
    W(t) & = & \dot{a}\rho - a\dot{\rho}\\
    G(t) & = & \int_{t^{'}}^{t} \frac{W^{2}}{2\rho^{2}} ds
\end{eqnarray*}
Letting
\begin{eqnarray*}
    Q(t) & = & \frac{x - a}{\rho}\\
    \mt & = & \int_{t^{'}}^{t} \rho^{-2} ds
\end{eqnarray*}
we have
\begin{eqnarray*}
    \int_{t^{'}}^{t^{''}}L_{0} dt & = & \int_{\mt^{'}}^{\mt^{''}}
        \bar{L}_{0} d\mt
\end{eqnarray*}
where
\begin{eqnarray*}
    \bar{L}_{0}\left(\bar{Q}(\mt),\dot{\bar{Q}}(\mt)\right) & = &
        \frac{1}{2}\left[\frac{d}{d\mt}\bar{Q}(\mt)\right]^{2} -
        \bar{F}\left(\bar{Q}(\mt)\right)
\end{eqnarray*}
Then our original propagator (\ref{2}) is given by
\begin{eqnarray}
\label{K}
    K\left(x^{''},t^{''};x^{'},t^{'}\right) & = &
        \left[\rho(t^{''})\rho(t^{'})\right]^{-1/2}e^{\frac{i}{\hbar}\left[X(t^{''})
        -
        X(t^{'})\right]}\bar{K}_{0}\left(\bar{Q}^{''},\mt^{''};\bar{Q}^{'},\mt^{'}\right)
\end{eqnarray}
where
\begin{eqnarray}
\label{Kbar}
    \bar{K}_{0}\left(\bar{Q}^{''},\mt^{''};\bar{Q}^{'},\mt^{'}\right)
        & = & \int
        e^{\frac{i}{\hbar}\int_{\mt^{'}}^{\mt^{''}}d\mt
        \bar{L}_{0}\left(\bar{Q}(\mt),\dot{\bar{Q}}(\mt) \right)}\mathcal{D}\left[\bar{Q}(\mt)\right]
\end{eqnarray}
See \cite{DhLa} for the measure factor
$\left[\rho(t^{''})\rho(t^{'})\right]^{-1/2}$ in (\ref{K}). Notice
that $\bar{K}_{0}$ is the "standard" path integral formula. We
will now rigorously define this using Muldowney's~\cite{Muld1}
application of the Henstock integral.

\section{Henstock Integration}
The Henstock (gauge or generalized Riemann) integral is a simple,
though powerful generalization of the Riemann integral.
Essentially it amounts to letting the interval length bounds
depend on position.  In the definition of the Riemann integral,
the bound is the same over the whole domain of integration.  Thus,
the Henstock integral allows us to take into account the local
variation of the functions we integrate.  Also, since it is a
non-absolute integral, it leads to a particularly simple and
conceptually pleasing definition of the Feynman path integral. For
more information on the Henstock integral in finite dimensions see
DePree and Swartz~\cite{DeSw}.  For the application to functional
integration see Muldowney~\cite{Muld2}.  The following is largely
based on \cite{Muld3} and \cite{Muld1}.

We begin with the Henstock integral in one dimension.  We do this
in $\Re$.  As shown in Example~(\ref{oneDint}) below, the
generalization to $E \subseteq \Re$ is easy. Let $I = [u,v), u,v
\in \bar{\Re} \doteq [-\infty,\infty]$ and $|I| = v-u$. If either
$u = -\infty$ or $v = \infty$ we let $|I| = 0$. Let $\delta (x) >
0$ for all $x \in  \bar{\Re}$.  We call $\delta$ a \textbf{gauge}.
We say that $I$ is \textbf{attached} to $x$ if any of the
following is true
\begin{eqnarray*}
I   =  \left\{
            \begin{array}{c}
                (-\infty,v) \\
                 \left[u,v\right)  \\
                  \left[u,\infty \right)    \\
            \end{array}
        \right\}
& and & x = \left\{
            \begin{array}{c}
                \mbox{$-\infty$}  \\
                \mbox{$u$ or $v$}  \\
                \mbox{$\infty$}
            \end{array}
            \right\}
\end{eqnarray*}
We call the attached pair $(x,I)$ $\delta$\textbf{-fine} if,
respectively
\[
\left\{
    \begin{array}{c}
        v < -\frac{1}{\delta (x)}  \\
        v-u < \delta (x) \\
        u > \frac{1}{\delta (x)}
    \end{array}
\right\}
\]
We say that
\begin{eqnarray*}
\mathcal{E} & = & \left\{ (x^{(i)},I^{(i)}) \right\}_{i = 1}^{n}
\end{eqnarray*}
is a \textbf{division} of $\Re$ if $I^{(i)} \cap I^{(j)} =
\emptyset$ for $i \neq j$ and
\begin{eqnarray*}
\bigcup_{i = 1}^{n} I^{(i)} & = & \Re
\end{eqnarray*}
Further, $\mathcal{E}_{\delta}$ is $\mathbf{\delta}$\textbf{-fine}
when every $(x^{(i)},I^{(i)})$ is $\delta$-fine for $i =
1,\ldots,n$.. When we sum over the point-interval pairs in
$\mathcal{E}_{\delta}$, we denote this by $\left(
\mathcal{E}_{\delta} \right) \Sigma$.
\begin{defi}
Let $h(x,I)$ be a function of point-interval pairs $(x,I)$ where
$h(x,I) \doteq 0$ when $x = \pm \infty$.  We say $h(x,I)$ is
\textbf{Henstock integrable over $\Re$}, to the value $\alpha$, if
for every $\epsilon > 0$ there exists a $\delta(x), x \in \Re$
such that
\begin{eqnarray*}
\left| \left( \mathcal{E}_{\delta} \right) \Sigma h(x,I) - \alpha
\right| & < & \epsilon
\end{eqnarray*}
whenever the division $\mathcal{E}_{\delta}$ is $\delta$-fine.  We
write
\begin{eqnarray*}
\int_{\Re} h(x,I) & = & \alpha
\end{eqnarray*}
\end{defi}
\begin{exam}
\label{oneDint}
Let
\begin{eqnarray*}
f(x) & = & \left\{ \begin{array}{ll}
            0 & \mbox{$x \in [0,1]$ and rational}   \\
            1 & \mbox{$x \in [0,1]$ and irrational}
            \end{array}
            \right.
\end{eqnarray*}
Here we'll let $h(x,I) = f(x)\left| I \right|$ where we'll define
the gauge below.  The gauge then puts an upper bound on the
possible size of the $\left| I \right|$'s.  Note that we do not
actually form an explicit set $\mathcal{E}_{\delta}$.  We have,
with our domain of integration being $[0,1]$
\begin{eqnarray*}
    \left| (\mathcal{E}_{\delta}) \sum_{j} f(x_{j})\left| I_{j} \right| -1
    \right| & = & \left| (\mathcal{E}_{\delta}) \sum_{j} \left(
    f(x_{j}) - 1 \right) \left| I_{j} \right| \right|
\end{eqnarray*}
If $x_{j}$ is irrational, $f(x_{j}) -1 = 0$.  We can let
$\delta(x) = 1$ for $x$ irrational.  For $x_{j} = q_{k}$ rational,
where $q_{k} \in \mathcal{Q} = \left\{q_{n} : q_{n}\ is\
irrational \right\}_{n=1}^{\infty}$, set $\delta(x_{j}) =
\frac{\epsilon}{2^{k+2}}$.  Then,
\begin{eqnarray*}
    \left| f(q_{k}) - 1 \right| \left| I_{k} \right| & < & \frac{\epsilon}{2^{k+1}}
\end{eqnarray*}
Since $q_{k}$ is attached to at most two intervals $I$, we have
(even if every rational number is attached to two intervals)
\begin{eqnarray*}
    \left| (\mathcal{E}_{\delta}) \sum_{j} \left(
    f(x_{j}) - 1 \right) \left| I_{j} \right| \right| & \leq & 2
    \sum_{k = 1}^{\infty} \frac{\epsilon}{2^{k+1}}  \\
    & = & \epsilon
\end{eqnarray*}
Hence,
\begin{eqnarray*}
    \int_{0}^{1} f(x) dx & = & 1
\end{eqnarray*}
\end{exam}

We'll now extend the Henstock integral to multiple dimensions. For
$\bar{\Re}^{2} = \bar{\Re}_{1} \times \bar{\Re}_{2}$ let $I
\subset \bar{\Re}^{2}$ be of the form
\begin{eqnarray*}
    I & = & I_{1} \times I_{2}      \\
        & = & [u_{1},v_{1}) \times [u_{2},v_{2})
\end{eqnarray*}
where $u_{i} = -\infty$ and $v_{i} = \infty$ are allowed.  All the
definitions are as in the one-dimensional case where now
$\mathbf{x}$ is attached to $I$ if $\mathbf{x}$ is at a corner of
$I$. With $\delta(\mathbf{x}) > 0$ for $\mathbf{x} \in
\bar{\Re}^{2}$ being our gauge, we say $(\mathbf{x},I)$ is
$\delta$-fine if $(x_{1},I_{1})$ and $(x_{2},I_{2})$ are both
$\delta$-fine.  Thus, the $I$'s will be rectangles in
$\bar{\Re}^{2}$.  Finally, $h(\mathbf{x},I)$ is integrable on
$\Re^{2}$, written as
\begin{eqnarray*}
    \int_{\Re^{2}} h(\mathbf{x},I) & = & \alpha
\end{eqnarray*}
if, for every $\epsilon > 0$ there exists a $\delta(\mathbf{x})$
such that
\begin{eqnarray*}
    \| \left(\mathcal{E}_{\delta} \right) \sum h(\mathbf{x},I) -
    \alpha \| & < & \epsilon
\end{eqnarray*}
whenever $\mathcal{E}_{\delta}$ is a $\delta$-fine division of
$\Re^{2}$.  The generalization to $\Re^{n}$ is obvious.

We now define the Henstock integral in infinite dimensions. Note
the definition of the gauge here.  This is different than that
defined in \cite{Muld1} (see \cite{Muld3}).  Let $B =
[t^{'},t^{''}]$, $x(t)$ be a function defined on $B$ and, for $N =
\{t_{i}\}_{i =1 }^{n - 1} \subseteq B$ let
\begin{eqnarray*}
I(N) & = & I_{t_{1}} \times \cdots \times I_{t_{n-1}} \subseteq
\Re^{n-1}
\end{eqnarray*}
where $I_{t_{i}} = [u_{i},v_{i}), (-\infty,v_{i})$ or
$[u_{i},\infty)$.  We call $N$ a \textbf{dimension set}.  $I(N)$
is \textbf{attached} to the vector $x(N) \in \Re^{n-1}$ when
\begin{eqnarray*}
x(N) & = & \left( x_{1},\ldots,x_{n-1} \right)
\end{eqnarray*}
where $x_{i} \doteq x(t_{i})$, $t_{i} \in N$ and $x_{i} = u_{i}$
or $v_{i}$,$-\infty$ or $\infty$, respectively.  Define
\begin{eqnarray*}
I[N] & \doteq & I(N) \times \Re^{B \setminus N}   \\
    & = & \left\{ x(t) : x \in \Re^{B} , x_{i} \in I_{t_{i}}
    \ for\ i=1,\ldots,n-1 \right\} \\
    & \doteq & I
\end{eqnarray*}
That is, $I[N]$ is the set of functions $x(t)$ defined on
$B=[t^{'},t^{''}]$ that pass through the interval $I_{t_{i}}$ at
time $t_{i}$.  We let
\begin{eqnarray*}
\left| I[N] \right| & \doteq & \left| I(N) \right|  \\
    & = & \left| I_{t_{1}} \right| \cdots \left| I_{t_{n-1}} \right|
    \\
    & = & Volume \left(I(N) \right)
\end{eqnarray*}
We say $x(t)$, $N$ and $I[N]$ are attached if $I(N)$ is attached
to $x(N)$.

For $M = \left\{t_{i} \right\} \subseteq B$ suppose we have $t^{'}
= t_{0} < t_{1} < \cdots < t_{m-1} < t_{m} = t^{''}$.  Let
\begin{eqnarray*}
J[M] & \doteq & \left\{ x(t) : x(t^{'}) = x^{'}, x(t^{''}) =
x^{''}, x_{i} \in I^{'}_{t_{i}} \ for\ i = 1,\ldots,m-1 \right\}
\end{eqnarray*}
$J[M]$ is the set of functions starting (ending) at $x^{'}$
($x^{''}$) and passing through the interval $I^{'}_{t_{i}}$ at
time $t_{i}$.  $J[M] \subseteq I[N]$ when $N \subseteq M$ or when
$I^{'}_{t_{i}} \subseteq I_{t_{i}}$ for $t_{i} \in M \cap N$.

Let $A = \left\{ t_{i} \right\}_{i=1}^{\infty} \subseteq B$.
Define $\mathcal{L}_{A}$ as the collection of all possible finite
subsets of $A$
\begin{eqnarray*}
\mathcal{L}_{A} & \doteq & \left\{ \left\{ {t_{j}^{k}} : t_{j}^{k}
\in A \ for\ j=1,\ldots,l_{k} \right\} \right\}_{k = 1}^{\infty}
\end{eqnarray*}
Let
\begin{eqnarray*}
L_{A}\left(x(t)\right): \bar{\Re}^{B} & \rightarrow &
\mathcal{L}_{A}
\end{eqnarray*}
and
\begin{eqnarray*}
    \bar{\delta}_{L_{A}} & \doteq & \left\{ \delta_{N}(x(N)) :
    L_{A}(x(t)) \subseteq N \ and\ x(N) \in \bar{\Re}^{n - 1} \right\}
\end{eqnarray*}
for $x(t) \in \bar{\Re}^{B}$.  Note that $\delta_{N}(x(N)) \in
\bar{\delta}_{L_{A}}$ is allowed to depend on both $x(N)$ and $N$.

A gauge is defined on $\bar{\Re}^{B}$ by
\begin{eqnarray*}
\gamma & = & \left( A , L_{A} , \bar{\delta}_{L_{A}} \right)
\end{eqnarray*}
Let $x(t)$, $N$, and $I[N]$ be attached.  Then
$\left(x(t),N,I\right)$ is $\gamma$\textbf{-fine} if $L_{A}
\subseteq N$ and $\left(x(N),I(N)\right)$ is $\delta_{N}$-fine on
the finite dimensional space $\Re^{N}$.  Here $\delta_{N} =
\delta_{N}\left(x(N)\right) \in \bar{\delta}_{L_{A}}$.

So, for each $x(t) \in \Re^{B}$ we have to sample the function at
the times $t_{i} \in L_{A}(x(t))$.  These times are constrained by
our choice of $A$.  However, we can sample the $x(t)$'s on sets
larger than the minimum, including $t_{i}$'s not in $A$.  We call
these larger sets $N \supseteq L_{A}(x(t))$.  By allowing the size
of $N$ to increase we can discriminate more among those functions
that give the same minimal set $L_{A}$.  The reason for this is
that these functions may take on different values at $t_{i}$ that
are not in $L_{A}$ even if they have the same values on $L_{A}$.
Now that we have a set N, we choose a finite dimensional gauge
$\delta_{N}$. This gauge gives us the size of the interval
associated with each $x(N) \in \bar{\Re}^{n-1}$.  We can shrink
this interval by choosing a finer gauge.

When $\left( x^{(i)}(t), I^{(i)}[N_{i}] \right)$ is $\gamma$-fine
for every $i$ we call
\begin{eqnarray*}
\mathcal{E}_{\gamma} & \doteq & \left\{ \left( x^{(i)}(t),
I^{(i)}[N_{i}] \right) \right\}_{i = 1}^{n}
\end{eqnarray*}
$\gamma$-fine.  $\mathcal{E}_{\gamma}$ is a division of $\Re^{B}$
when
\begin{eqnarray*}
I^{(i)}[N_{i}] \cap I^{(j)}[N_{j}] & = & \emptyset
\end{eqnarray*}
for $i \neq j$ and
\begin{eqnarray*}
\bigcup_{i = 1}^{n} I^{(i)}[N_{i}] & = & \Re^{B}
\end{eqnarray*}
\begin{defi}
\label{infdint}
Let $h(x,N,I)$ be a function of point-dimension
set-interval triplets.  Define $h(x,N,I) \doteq 0$ whenever
$x(t_{i}) = \pm \infty$ for $t_{i} \in N$.  We say that $h(x,N,I)$
is \textbf{Henstock integrable over $\Re^{B}$}, to the value
$\alpha$, if for every $\epsilon > 0$ there exists a $\gamma$ such
that
\begin{eqnarray*}
\left| \left( \mathcal{E}_{\gamma} \right) \Sigma h(x,N,I) -
\alpha \right| & < & \epsilon
\end{eqnarray*}
whenever the division $\mathcal{E}_{\gamma}$ is $\gamma$-fine.  We
write
\begin{eqnarray*}
\int_{\Re^{B}} h(x,N,I) & = & \alpha
\end{eqnarray*}
\end{defi}

\section{Feynman Path Integration}

Our path integral is given by (\ref{Kbar})
\begin{eqnarray*}
    \bar{K}_{0}\left(\bar{Q}^{''},\mt^{''};\bar{Q}^{'},\mt^{'}\right)
        & = & \int
        e^{\frac{i}{\hbar}\int_{\mt^{'}}^{\mt^{''}}d\mt
        \bar{L}_{0}\left(\bar{Q}(\mt),\dot{\bar{Q}}(\mt) \right)}\mathcal{D}\left[\bar{Q}(\mt)\right]
\end{eqnarray*}
We start with a finite dimensional approximation to this
\begin{eqnarray*}
    \bar{K}_{0} & \approx & \int_{u_{1}}^{v_{1}} \cdots
    \int_{u_{n-1}}^{v_{n-1}} exp \left[ \frac{i}{\hbar} \sum_{j =
    1}^{n} \left\{ \frac{(\bar{Q}_{j} -
    \bar{Q}_{j-1})^{2}}{2(\mt_{j} - \mt_{j-1})} -
    \bar{F}\left(\bar{Q}_{j-1} \right)(\mt_{j} - \mt_{j-1}) \right\}
    \right] \times      \\
     & & \prod_{j = 1}^{n} \left[ 2 \hbar \pi i (\mt_{j} -
    \mt_{j-1}) \right]^{-1/2} d\bar{Q}_{1} \cdots d\bar{Q}_{n-1}
\end{eqnarray*}
where $\bar{Q}_{j} \doteq \bar{Q}(\mt_{j})$.  Define
\begin{eqnarray*}
    g_{\bar{F}}(\bar{Q},N) & \doteq & exp \left[ \frac{i}{\hbar}
    \sum_{j =
    1}^{n} \left\{ \frac{(\bar{Q}_{j} -
    \bar{Q}_{j-1})^{2}}{2(\mt_{j} - \mt_{j-1})} -
    \bar{F}\left(\bar{Q}_{j-1} \right)(\mt_{j} - \mt_{j-1}) \right\}
    \right] \times      \\
     & & \prod_{j = 1}^{n} \left[ 2 \hbar \pi i (\mt_{j} -
    \mt_{j-1}) \right]^{-1/2}
\end{eqnarray*}
where
\begin{eqnarray*}
g_{0}(\bar{Q},N) & = & exp \left[ \frac{i}{\hbar} \sum_{j =
    1}^{n} \left\{ \frac{(\bar{Q}_{j} -
    \bar{Q}_{j-1})^{2}}{2(\mt_{j} - \mt_{j-1})} \right\} \right]
     \prod_{j = 1}^{n} \left[ 2 \hbar \pi i (\mt_{j} -
    \mt_{j-1}) \right]^{-1/2}
\end{eqnarray*}
is the free particle case.  Note that the above can be interpreted
either as integrating over functions or over the finite
dimensional space $\Re^{n-1}$.  If we wish to specify the latter
case we will write $g_{\bar{F}}(\bar{Q}(N))$ instead.

Now define
\begin{eqnarray*}
    g_{\bar{F}}(\bar{Q},N,I) & \doteq & g_{\bar{F}}(\bar{Q},N) \left| I[N] \right|  \\
    g_{\bar{F}}(\bar{Q}(N),I(N)) & \doteq & g_{\bar{F}}(\bar{Q}(N)) \left| I(N) \right|
\end{eqnarray*}
and, having chosen a $J[N]$ for a fixed $N$, let
\begin{eqnarray*}
    G_{\bar{F}}(J(N)) & \doteq & \int_{J(N)} g_{\bar{F}}(\bar{Q}(N),I(N))
\end{eqnarray*}
This is a finite dimensional integral over a hypercube in
$\Re^{n-1}$ given by
\[
    I_{t_{1}}^{'} \times \cdots \times I_{t_{n-1}}^{'}
\]
where $\bar{Q}_{i} \in I_{t_{i}}^{'}$ for $\bar{Q}(t) \in J[N]$.
We then let $G_{\bar{F}}(J[N]) = G_{\bar{F}}(J(N))$.  In
particular we have
\begin{eqnarray*}
    G_{0}(J) & = & \int_{J} g_{0}(\bar{Q}(N),I(N))    \\
        & = & \int_{J} g_{0}(\bar{Q}(N)) \left| I(N) \right|   \\
        & = & \int_{J} exp \left[ \frac{i}{\hbar} \sum_{j =
    1}^{n} \left\{ \frac{(\bar{Q}_{j} -
    \bar{Q}_{j-1})^{2}}{2(\mt_{j} - \mt_{j-1})} \right\} \right]
     \prod_{j = 1}^{n} \left[ 2 \hbar \pi i (\mt_{j} -
    \mt_{j-1}) \right]^{-1/2} \left| I(N) \right|
\end{eqnarray*}
for the free particle case.  Let
\begin{eqnarray*}
    \bar{f}(\bar{Q},N) & \doteq & exp \left[ - \frac{i}{\hbar} \sum_{j =
    1}^{n} \bar{F}  \left( \bar{Q}_{j-1} \right) (\mt_{j} - \mt_{j-1})
    \right]
\end{eqnarray*}
Then, if $\bar{F}$ is continuous we have
\begin{eqnarray}
\label{Prop}
\bar{K}_{0}(\bar{Q}^{''},\mt^{''};\bar{Q}^{'},\mt^{'}) & = &
    \int_{\Re^{(\mt^{'},\mt^{''})}} \bar{f}(\bar{Q},N)G_{0}(I)
\end{eqnarray}
Letting $h( \bar{Q},N,I ) = \bar{f}(\bar{Q},N)G_{0}(I)$ this
integral is defined  in Definition (\ref{infdint}).

We would like a better way to evaluate (\ref{Prop}).  To that end
we let $\tau_{j} = \mt^{'} + j(\mt^{''}-\mt^{'})/m$ for $j =
1,\ldots,m-1$ and $m=2^{q}$.  Then we have $M = \left\{ \tau_{j}
\right\}_{j=1}^{m-1}$ and, with $y_{j} \doteq y(\tau_{j})$,
\[
    y  =  \left(y_{1},\ldots,y_{m-1}\right) \in \Re^{m-1}
\]
where $y_{0} = \bar{Q}^{'}$, $y_{m} = \bar{Q}^{''}$.  Letting
\begin{eqnarray*}
    g_{\bar{F}}^{(m)}(y) & \doteq & exp \left[ \frac{i}{\hbar} \sum_{j
    =
    1}^{m} \left\{ \frac{(y_{j} - y_{j-1})^{2}}{2(\tau_{j} - \tau_{j-1})} -
    \bar{F}\left(y_{j-1} \right)(\tau_{j} - \tau_{j-1}) \right\}
    \right] \times      \\
     & & \prod_{j = 1}^{m} \left[ 2 \hbar \pi i (\tau_{j} -
    \tau_{j-1}) \right]^{-1/2}
\end{eqnarray*}
we have that if
\begin{eqnarray*}
    \int_{\Re^{m-1}} g_{\bar{F}}^{(m)}(y) dy & = & \Gamma_{m}
\end{eqnarray*}
then
\begin{eqnarray*}
    \int_{\Re^{(\mt^{'},\mt^{''})}}\bar{f}(\bar{Q},M)G_{0}(I) & = & \Gamma_{m}
\end{eqnarray*}
We'll now show that Feynman's limit
\[
    \lim_{m \rightarrow \infty} \int_{\Re^{m-1}} g_{\bar{F}}^{(m)}(y) dy
\]
is in fact the correct method for determining the propagator.
First, we need the following definition.
\begin{defi}
Let $h_{m}(y^{(m)})$ be defined for $y^{(m)} \in \Re^{m}$ and, for
every $\epsilon > 0$ let there exist positive functions $\left\{
\delta_{0}^{(m)} \right\}$ such that $\delta_{0}^{(m)}(y^{(m)}) =
\delta_{0}^{(m+1)}(y^{(m+1)})$.  Then $h_{m}$ is \textbf{uniformly
integrable} if for all $\delta_{0}^{(m)}$-fine divisions of
$\Re^{m}$ we have
\begin{eqnarray*}
    \left| (\mathcal{E}_{\delta_{0}^{(m)}}) \sum h_{m}(y^{(m)}) \left|
    I(M) \right| - H_{m} \right| & < & \varepsilon
\end{eqnarray*}
\end{defi}
If $g_{\bar{F}}^{(m)}(y)$ is uniformly integrable and $\Gamma_{m}
\rightarrow \Gamma$ as $m \rightarrow \infty$ then

\begin{eqnarray*}
    \lim_{m \rightarrow \infty} \int_{\Re^{m-1}} g_{\bar{F}}^{(m)}(y) dy
    & = & \bar{K}_{0}\left( \bar{Q}^{''},\mt^{''}; \bar{Q}^{'},\mt^{'} \right)
\end{eqnarray*}

\section{Evolutionary Semigroups}
In this section we develop some evolution semigroup theory.  For a
more detailed coverage of this see Chicone and
Latushkin~\cite{ChLa}.  For the use of evolution semigroups in
control theory see Curtain and Zwart~\cite{CuZw}, Clark, et.
al~\cite{ClLa} and Bensoussan, et. al~\cite{BeDa1,BeDa2}.


  \subsection{Evolutionary Families}
  Our goal is to be able to deal with the \textit{non-autonomous abstract Cauchy
  problem} in a
  Hilbert space $\mathcal{H}$
  \begin{eqnarray}
  \label{nacp}
  \dot{u}(\theta) & = & A(\theta)u(\theta) \\
  u(\tau) & = & x_{\tau} \nonumber
  \end{eqnarray}
  where $\theta \geq \tau \in \Re_{+}$ and $x_{\tau} \in D(A(\tau))$
   is the initial condition that must lie in the domain of $A$, $D(A(\cdot))$, at time $\tau$.
  When we think about solutions to (\ref{nacp}) the following
  concept proves useful.
  \begin{defi} A \textbf{strongly continuous evolutionary
  family on $\Re_{+}$} of bounded
  operators $U(\theta,\tau)$ on a Hilbert space $\mathcal{H}$ has
  the following properties
  \begin{itemize}
  \item \(U(\theta,\theta) = I$ for $\theta \in \Re_{+}\).
  \item \(U(\theta,\tau) = U(\theta,r)U(r,\tau)\) for \(\theta \geq r\geq \tau \in \Re_{+}\).
  \item The map \((\theta,\tau) \rightarrow U(\theta,\tau)\) is strongly continuous for
  \(\theta \geq \tau \in \Re_{+}\).
  \end{itemize}
  \end{defi}
  For a given \U, its \textbf{growth bound}
  is given by
  \[
  \omega_{0}(U) \doteq inf \{\omega\in \Re:\exists
  M_{\omega}\geq 1\ where\ \|\U\|\leq M_{\omega}e^{\omega (\theta - \tau)}
  \ \forall\ \theta \geq \tau\}
  \]
  If $\omega_{0}(U) < 0$ we call \U\ \textit{exponentially
  stable} and if $\omega_{0}(U) <
  \infty$ we call \U\ \textit{exponentially
  bounded}.  It is fairly easy to see that our propagator
  $K\left(x^{''},t^{''};x^{'},t^{'}\right)$ gives rise to an exponentially
  bounded, strongly continuous evolutionary family through the
  operator
  \begin{eqnarray*}
    U_{K}^{(x^{''})}(\theta,\tau)\psi_{\tau}(x^{'}) & = & \int dx^{'} K\left(
    x^{''},\theta;x^{'},\tau \right)\psi_{\tau}(x^{'})
  \end{eqnarray*}

  \subsection{Howland Semigroups on $\Re_{+}$}
  Now, if \U\ acts on the Hilbert space $\mathcal{H}$, we will
  let $\mathbf{H}_{+} \doteq L_{p}(\Re_{+},\mathcal{H}),\ 1
  \leq p < \infty$ where for $h \in \mathbf{H}_{+}$ we have
  $h:\Re_{+} \rightarrow \mathcal{H}$.  If, for example,
  $\mathcal{H}$ is $L_{p}(\Re_{+})$ then $\mathbf{H}_{+} =
  L_{p}(\Re_{+},L_{p}(\Re_{+})) =
  L_{p}(\Re_{+} \times \Re_{+})$ and, for $h \in
  \mathbf{H}_{+},\ h:\Re_{+} \rightarrow
  L_{p}(\Re_{+})$ or $\tilde{h}(\Re_{+} \times
  \Re_{+}) \in L_{p}(\Re_{+} \times
  \Re_{+})$ where $\tilde{h}(x, y = a) \doteq
  \left.h(x)\right|_{y = a}$.
  \begin{defi}
  \label{Howsemi}
  If $\U,\ \theta \geq \tau \geq 0$, is a strongly
  continuous exponentially bounded evolution family on Hilbert
  space $\mathcal{H}$, the \textbf{associated evolutionary
  semigroup}~\index{associated evolution semigroup} $E^{t}_{+},\ t
  \geq 0$, acting on $\mathbf{H}_{+} =
  L_{p}(\Re_{+},\mathcal{H})$ is given by
  \begin{eqnarray}
  \label{howl}
  (\Ep h)(\theta) \doteq \left\{ \begin{array} {ll}
                                U(\theta,\theta - t)h(\theta - t)
                                & \theta \geq t \geq 0 \\
                                0 & 0 \leq \theta \leq t
                                \end{array} \right.
  \end{eqnarray}
  We also call (\ref{howl}) a \textbf{Howland semigroup on the half line
  ($\Re_{+}$)}.
  \end{defi}
  For our propagator, the associated evolutionary semigroup is given by
  \begin{eqnarray*}
  (\Ep \psi)(\theta) & \doteq & \left\{ \begin{array} {ll}
                                \int K(x^{''},\theta ; x^{'},\theta - t)
                                \psi(x^{'},\theta - t) dx^{'}
                                & \theta \geq t \geq 0 \\
                                0 & 0 \leq \theta \leq t
                                \end{array} \right.
  \end{eqnarray*}
  We have the following.
  \begin{prop}\label{evsemipr} \Ep\ as defined in (\ref{howl}) is a strongly
  continuous semigroup.\\ \\
  \verb"PROOF."  We first show that \Ep\ is a semigroup.  We have
  \begin{eqnarray*}
  (\Ep h)(\theta) & = & U(\theta,\theta - t)h(\theta - t)   \\
  & \doteq & g(\theta)
  \end{eqnarray*}
  for $\theta \geq t \geq 0$.  Then
  \begin{eqnarray*}
  (E^{s}_{+}g)(\theta) & = & U(\theta,\theta - s)g(\theta - s)   \\
  & = & U(\theta,\theta - s)U(\theta - s,\theta - s - t)h(\theta - s - t)    \\
  & = & U(\theta,\theta - s - t)h(\theta - s - t)  \\
  & = & (E^{s + t}_{+}h)(\theta)
  \end{eqnarray*}
  Also,
  \begin{eqnarray*}(E^{0}_{+}h)(\theta) & = & U(\theta,\theta)h(\theta)   \\
  & = & Ih(\theta)
  \end{eqnarray*}
  Then, $\Ep E^{s}_{+} = E^{s+t}_{+}$ and $E^{0}_{+} = I$.  So \Ep\ is a
  semigroup.

  Now we show strong continuity of \Ep.  First, the set of
  compactly supported, continuous functions
  \(C_{c}(\Re,\mathcal{H})\) is dense in $L_{p}$.  We have for
  \(h \in C_{c}(\Re,\mathcal{H})\)
  \[
  \lim_{t \rightarrow 0^{+}} \Ep h = h
  \]
  Also, we can find a \(\delta > 0\) and an \( N \geq 1\) such
  that, for all \( t \in [0,\delta]\)
  \begin{eqnarray*}
  \|\Ep\| & \leq & Me^{\omega t}   \\
  & \leq & N
  \end{eqnarray*}
  This follows from the exponential boundedness of \U as given in Definition~(\ref{Howsemi}).  These two
  conditions imply strong continuity~\cite[page 38]{EnNa}.
  Hence, \Ep\ is a strongly continuous semigroup on
  $L_{p}(\Re_{+},\mathcal{H})$ where $1 \leq p < \infty$ and $t \geq
  0$.$\spadesuit$
  \end{prop}

  Consider the \textit{right translation semigroup}~\index{right translation semigroup}
  on $L_{p}(\Re_{+}),\ 1 \leq p < \infty$
  \begin{eqnarray*}
  (T_{r}(t)f)(\theta) & \doteq & \left\{ \begin{array}{ll}
                                    f(\theta - t) & ,\theta - t
                                    \geq 0  \\
                                    0 & ,\theta - t \leq 0
                                    \end{array} \right.
  \end{eqnarray*}
  We have that
  \begin{eqnarray*}
  A_{r}f & \doteq & \lim_{h \rightarrow 0^{+}} \frac{T_{r}(h)f - f}{h}
  \\
  & = & \lim_{h \rightarrow 0^{+}} \frac{f(\theta - h) -
  f(\theta)}{h} \\
  & = & -\frac{d}{d\theta} f(\theta)
  \end{eqnarray*}
  So, the generator of $T_{r}(t)$ is $A_{r} = -\frac{d}{d\theta}$.
  We have~\cite[pages 66-67]{EnNa}
  \[
  D(A_{r}) = \left\{f\in\Lp:f\ is\ absolutely\ continuous\ and\ f^{'}\in\Lp\right\}
  \]
  when $\mathcal{H} = \Lp,\ 1 \leq p < \infty$.  With
  \((T_{r}(t)h)(\theta) = h(\theta - t)\) and defining the weight
  $w(\theta;t) \doteq U(\theta,\theta - t)$ we have that $\Ep =
  w(\theta;t)T_{r}(t)$ is a weighted translation operator.

\section{Acknowledgements}

We would like to thank Pat Muldowney for his valuable information
on the Henstock integral, especially for reference~\cite{Muld3}.
Thanks also to Richard Eykholt for reading and commenting on the
draft. As always, the mistakes are the authors' own.

\bibliographystyle{plain}
\bibliography{CT01}

\end{document}